\documentclass[12pt]{iopart}

\usepackage{graphicx}
\usepackage{iopams}

\newcommand{\R}{\ensuremath{\mathbb{R}}}
\newcommand{\Half}{\ensuremath{\mathbb{H}}}
\newcommand{\SLEk}{$SLE_{\kappa}$ }
\newcommand{\SLEkp}{$SLE(\kappa,\rho)$ }
 
\begin{document}
 
\title{\SLEkp processes, hiding exponents and self-avoiding walks in a wedge}

\author{Nathan Deutscher$^1$ and Murray T. Batchelor$^{1,2}$}
\address{$^1$ %Department of Mathematics, 
Mathematical Sciences Institute,  Australian National University, \\
Canberra ACT 0200, Australia}
\address{$^2$ Department of Theoretical Physics, 
Research School of Physical Sciences and Engineering, 
Australian National University, Canberra ACT 0200, Australia}
\ead{Murray.Batchelor@anu.edu.au}

\begin{abstract}
This article employs Schramm-Loewner Evolution to obtain intersection exponents for 
several chordal $SLE_{8/3}$ curves in a wedge. 
As $SLE_{8/3}$ is believed to describe the continuum limit of self-avoiding walks, 
these exponents correspond to those obtained by Cardy, Duplantier and Saleur for 
self-avoiding walks in an arbitrary wedge-shaped geometry using conformal invariance  based arguments.
Our approach builds on work by Werner, where the restriction property for \SLEkp processes and an 
absolute continuity relation allow the calculation of such exponents in the half-plane. 
Furthermore, the method by which these results are extended is general enough to apply to the 
new class of hiding exponents introduced by Werner.
\end{abstract}

%Random walks and Levy flights
%\pacs{05.40.Fb}
%Random walks, random surfaces, lattice animals, etc.
%\ams{82B41}
%Comment to leave out Submitted to journal title message (don't seem to have this journal in abbreviations bit!)
%\submitto{J. Phys. A: Math. Theor.}

\section{Introduction}

Schramm-Loewner Evolution ($SLE$) processes have proven an invaluable tool in investigating the 
continuum limit of random curves \cite{KN04,Law05,Car05,BB06}. 
In particular, the $SLE$ formalism has provided rigorous proofs of previously established results, 
such as Cardy's formula for crossing probabilities between segments of the boundary of a 
compact two-dimensional region at the percolation threshold \cite{Car92}, 
as well as numerous new results on problems that had previously eluded concrete analysis. 
Another early success of the $SLE$ approach was the calculation of
 intersection exponents between Brownian motions 
in whole and half-plane geometries \cite{LSW01a,LSW01b}. 
Here we consider intersection exponents in a wedge-shaped geometry of opening angle $\theta \pi$.

The first derivation of intersection exponents between Brownian motions drew on a special case of $SLE$
in which an additional property holds, namely the locality of $SLE_{6}$. 
Similarly a not unrelated restriction property holds for $SLE_{8/3}$ and enhances the ability to calculate 
certain probabilities. 
In addition, as the only \SLEk process to satisfy the restriction property, $SLE_{8/3}$ is the only possible 
conformally invariant continuum limit for the self-avoiding walk. 
Although the existence and conformal invariance of such a limit is yet to be proven, 
the link between $SLE_{8/3}$ and the self-avoiding walk has been fleshed out in Ref.~\cite{LSW04}, 
and corresponding predictions numerically confirmed \cite{Ken02, Ken04}.

Boundaries of other sets satisfying the restriction property can be constructed using the generalisation 
of \SLEk to an \SLEkp process, as detailed in Ref.~\cite{LSW03}. 
An \SLEkp process may be pictured as an \SLEk curve with a drift dependent on the $\rho$ parameter. 
Relatively recently, additional absolute continuity relations between \SLEkp processes have been 
established by Werner \cite{Wer04}. 
This is a particularly powerful result, as it allows us to get a handle on mutually avoiding curves, 
something standard \SLEk techniques are troubled by. 
Alternative methods of incorporating mutual avoidance into \SLEk involve ideas originating in 
quantum gravity \cite{Dup06}.

With these properties of \SLEkp established, Werner was able to calculate intersection exponents 
for several $SLE_{8/3}$ in the half-plane, corresponding to previous exponents obtained for the 
self-avoiding walk \cite{Car84,CR84,DS86}. 
In addition Werner calculated a new class of exponents, not found in the physics literature, 
which he termed hiding exponents. 
In this paper we extend both sets of exponents to wedge geometries. 
This yields the counting exponents for several self-avoiding walks (stars) in a wedge, as determined 
previously \cite{DS86}. 
In using \SLEkp techniques we ensure that this derivation is in fact complete, modulo the assumption that 
$SLE_{8/3}$ is indeed the scaling limit of the self-avoiding random walk. 
We also extend Werner's hiding exponents \cite{Wer04}, indicating the generality of this approach.

The outline of the paper is as follows. In Section 2 we first briefly review the results of 
Refs.~\cite{LSW03} and \cite{Wer04} on \SLEkp processes.
We then show in Section 3 how these results are used to obtain intersection and hiding exponents 
in the half plane. 
In Section 4 we show how the restriction property allows a neat calculation to transfer these results 
across into the wedge geometry, and discuss these results in terms of self-avoiding walks.
Concluding remarks are given in Section 5.

\section{\SLEkp processes and their properties}

In this section we recall the definition of \SLEkp and draw upon past results concerning its properties. 
The first results relate \SLEkp to the boundary of one-sided restriction measure samples. 
The second then establish that the law of an \SLEkp conditioned not to intersect such a boundary is 
itself an \SLEkp with a perturbed parameter $\rho$. 
It is not difficult to see that these twin results may provide powerful iterative techniques for investigating 
mutually avoiding interfaces.

\subsection{\SLEkp processes}

First, recall the definition of a standard \SLEk process. 
The family of conformal maps $(g_{t},t\geq0)$ associated with such a process are the solutions to the 
chordal Loewner equation
\begin{equation}
\label{cle}
\partial_{t} g_{t}(z) = \frac{2}{g_{t}(z)-W_{t}},
\end{equation}
with driving function $W_{t}$ simply a scaled Brownian motion; $W_{t}:=\sqrt{\kappa}B_{t}$. 
At each time $t$, this gives rise to a conformal map $g_{t}$ from a domain $H_{t}$ onto $\Half$, 
where we may define $H_{t}=\{z:|g_{s}(z)-W_{s}|>0, \forall s \in [0,t]\}$. In particular these maps $g_{t}$ 
define a family of growing subsets $K_{t}:=\Half\backslash H_{t}$ of the complex half-plane, 
which we may think of as being generated by a path (this happens with probability 1 \cite{RS05}). 
This path is permitted to reflect off itself and the real line, and is often itself referred to as an 
\SLEk  process.
One result regarding this path is its dimension, established with proof in Ref. \cite{B03},
\begin{equation}
\label{sledim}
d_{\kappa} = \max\left\{1+{\kappa}/{8}\right\}.
\end{equation}

The generalisation of \SLEk involves adding a drift term to the driving function. 
We envisage this as equivalent to adding a pressure on the left side of the \SLEk path
that pushes it in a particular direction. 
To be precise we take $\rho>-2$ and let
\begin{eqnarray}
W_{t}	&	=	&	\sqrt{\kappa}B_{t}+\int_{0}^{t} \frac{\rho}{W_{s}-O_{s}} ds, \\
O_{t}		&	=	&	\int_{0}^{t} \frac{2}{W_{s}-O_{s}} ds,
\end{eqnarray}
and call the solution $(g_{t},t\geq0)$ to (\ref{cle}) with this driving function $SLE(\kappa,\rho)$.
Note that if $\rho$ is set equal to zero we return to a standard $SLE_{\kappa}$. 
Suppose that the Brownian motion is begun at a point $a$ on the real line. 
Then $(O_{0},W_{0})=(0,a)$ and we say that the \SLEkp process is started from this pair of points.

An alternate way of constructing the pair $(O_{t},W_{t})$ begins by defining $Y_{t}$, 
a $d$-dimensional Bessel process where
\begin{equation}
\label{dimension}
d=1+{2(\rho+2)}/{\kappa}.
\end{equation} 
The $\rho>-2$ restriction stems from this association. 
In addition, it can be shown that by taking $d\geq2$ we ensure that the \SLEkp curve never 
hits the real axis to the left of its starting point. 
More importantly, it is this perspective on the driving function that allowed Werner to establish 
an absolute continuity relation between \SLEkp processes. 
Before turning to this, we discuss the context in which \SLEkp was first introduced, that of the restriction property.

\subsection{\SLEkp and the restriction property}

The \SLEk approach is at its most powerful when coupled with additional properties. 
One of these is the restriction property. 
This was first formalised in Ref.~\cite{LSW03} and it is this that motivated the extension to $SLE(\kappa,\rho)$.

We begin by stating what is meant by one-sided restriction.
First, let $\mathcal{A}$ be the set of all closed subsets $A\subset\overline{\Half}$ such that 
\begin{itemize}
	\item $\Half\backslash A$ is simply connected.
	\item $A$ is bounded and bounded away from the negative reals.
\end{itemize}
To each $A\in\mathcal{A}$ we associate a unique conformal map $\Phi_{A}$ that maps $\Half\backslash A$ onto $\Half$. Uniqueness is obtained by forcing $\Phi_{A}$ to fix $0$ and $\infty$ and asking that $\Phi_{A}(z)/z\rightarrow 1$ 
as $z\rightarrow\infty$. 
Second, a closed subset $K\subset\overline{\Half}$ is \textit{left-filled} if $K\cap \R=(-\infty,0]$ and both $K$ and 
$\Half\backslash K$ are unbounded and simply connected.

Finally, we say a random left-filled set satisfies \textit{one-sided restriction} if for all $A\in\mathcal{A}$ the 
law of $K$ is identical to the law of $\Phi_{A}(K)$ conditioned on the event $\{K\cap A = \emptyset \}$. 
It can be shown \cite{LSW03} that this implies the existence of a positive number $\alpha$ 
such that for all $A\in\mathcal{A}$
\begin{equation}
\label{restriction}
\textbf{P}[K \cap A=\emptyset] = \Phi_{A}'(0)^{\alpha}.
\end{equation}
This is a powerful result, and the one which will enable us to extend half-plane exponents to their analogues in a wedge. 
We note that the converse to (\ref{restriction}) has also been discussed \cite{LSW03}, 
with the conclusion that for each $\alpha>0$ there exists a unique random left-filled set such that (\ref{restriction}) is satisfied. 
The law of such a set is called the one-sided restriction measure of exponent $\alpha$. 
It is shown \cite{LSW03} that the boundary of a sampled one-sided restriction measure of exponent $\alpha$ is an 
$SLE(8/3,\rho)$ process where
\begin{equation}
\label{SLEkpREST}
\alpha = \frac{1}{32}{(\rho+2)(3\rho+10)}.
\end{equation}
This result has been extended in \cite{Dub05} to cases of $\kappa\not= 8/3$. 
We will not be considering such cases in this paper, although the extension to these given the method 
we detail would be straightforward. 
It is also worth pointing out that a Brownian motion conditioned to stay in the half-plane is a 
restriction measure with exponent $1$. 
This gives a way in which to picture arbitrary restriction measure samples of exponent $\alpha$ 
as simply a collection of $\alpha$ Brownian motions. 

On a final note, the result (\ref{SLEkpREST}) shows that $SLE_{8/3}$ satisfies the one-sided restriction property 
(and more generally the concept of two-sided restriction, see \cite{LSW03}) with exponent $5/8$. 
It was this observation that led to the conjecture that the scaling limit of the self-avoiding walk in the half-plane is $SLE_{8/3}$. 
This conjecture has been further fleshed out in Ref.~\cite{LSW04}, and has received strong support in numerical 
tests by Kennedy \cite{Ken02, Ken04}.

\subsection{Absolute continuity relations}

The second of the two properties is a little more involved in its set up and we refer the 
interested reader to Ref.~\cite{Wer04} for details. 
Essentially, absolute continuity results between Bessel processes of different dimensions 
$d$ follow from Girsanov's transformation and translate into analagous results for \SLEkp for differing 
$\rho$ (see equation (\ref{dimension})). 
The final outcome is that an \SLEkp process conditioned to avoid a one-sided restriction measure of 
exponent $\alpha$ is itself an $SLE(\kappa,\bar{\rho})$ process with
\begin{equation}
\label{SLEkpAVOID}
\bar{\rho} = \frac12 {\kappa}-2 + \kappa \sqrt{\frac{4\alpha}{\kappa}+\left(\frac{\rho+2}{\kappa}-\frac{1}{2}\right)^{2}}. 
\end{equation}
Also, an \SLEkp process started at a point $a>0$, and run up until time 1, will intersect a one-sided restriction 
sample of exponent $\alpha$ with a probability that decays like $a^{\sigma}$ as $a\to 0$ where
\begin{equation}
\label{SLEkpEXPO}
\sigma = - \left(\frac{\rho+2}{\kappa}-\frac{1}{2}\right) + 
\sqrt{\frac{4\alpha}{\kappa}+\left(\frac{\rho+2}{\kappa}-\frac{1}{2}\right)^{2}}. 
\end{equation}

%Exponents in the Half-Plane

\section{Exponents in the half-plane}

The twinned properties of restriction and the absolute continuity relation are now used to introduce  
exponents calculated by Werner \cite{Wer04} which we soon extend to wedge geometries. 
We begin with the new class of hiding exponents introduced by Werner.

\subsection{Hiding exponents}

%%%%%%%%%%%%%%%%%%%%%%%%%%%%%%%%%%%%%%%%%%%%%%%%%%%%%%%%%%%%
\begin{figure}
  \centering
  \scalebox{0.22}{\includegraphics{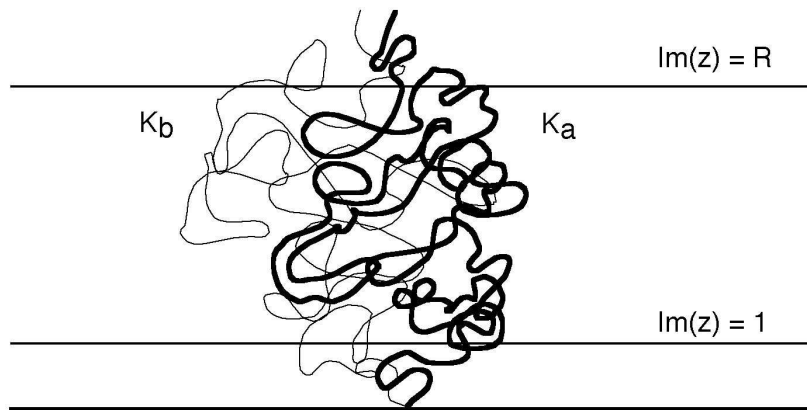}}
  \caption{Two independent samples of one-sided restriction measures, with $K_{a}$ `hiding' $K_{b}$ on the right. 
  Note that $K_{b}$ does not `hide' $K_{a}$ on the left, although these `two-sided' hiding exponents have been  
  discussed \cite{Wer04}. For simplicity $K_{a}$ and $K_{b}$ are drawn as the union of $a$ and $b$ 
  independent Brownian motions in the half-plane (here $a=b=1$); the true restriction samples are the associated left-filled sets.}
   \label{hiding}
\end{figure}
%%%%%%%%%%%%%%%%%%%%%%%%%%%%%%%%%%%%%%%%%%%%%%%%%%%%%%%%%%%%

The first exponent is almost immediate from that of the last section. 
An $SLE(8/3,\rho)$ process started at a point $a>0$ and run up until time 1 is itself the right boundary of a one-sided restriction sample of exponent $\alpha$. 
This exponent can be calculated from the formula (\ref{SLEkpREST}), which we invert to give
\begin{equation}
\label{he1}
\rho = \frac13(-8+2\sqrt{1+24\alpha}),
\end{equation}
with the other root impossible as $\rho>-2$. 
We require also that our $SLE(8/3,\rho)$ process avoids the negative real axis, so that the dimension $d$ 
from (\ref{dimension}) is $d\geq 2$, and hence $\rho\geq -2/3$ implying $\alpha\geq 1/3$. 
Now our $SLE(8/3,\rho)$ process (started at $a$, run to time 1), is the right boundary of a one-sided restriction 
measure sample of exponent $\alpha$, and avoids a second one-sided restriction measure sample 
of exponent $\beta$ with a probability that decays like $a^{\sigma}$ where $\sigma$ was as given in (\ref{SLEkpEXPO}). 
Substituting (\ref{he1}) into (\ref{SLEkpEXPO}) we obtain
\begin{equation}
\label{he2}
\sigma = \frac{1}{4}\left(-3 - \sqrt{1+24\alpha}+ \sqrt{24\beta+(\sqrt{1+24\alpha}-3)^{2}}\right).
\end{equation}

This exponent has been constructed to describe the decay in the probability that one sample of a restriction measure avoids the right boundary of another: that is, the second sample hides the first from one side of the half-plane. 
To be explicit, consider independent one-sided restriction measure samples $K_{\alpha}$ and $K_{\beta}$ indexed by their exponents. 
Then the probability that the right boundary of $K_{\alpha} \cup K_{\beta}$ in the strip $\{z:1 \geq \Im(z) \leq R\}$ contains no points in $K_{\beta}$ decays like $R^{-\sigma}$ as $R\rightarrow\infty$ where $\sigma$ is as in equation (\ref{he2}). 
This scenario is illustrated in Figure \ref{hiding}.

As a special case of the last, let $\alpha=\beta=5/8$. 
In this case both $K_{\alpha}$ and $K_{\beta}$ are simple independent $SLE_{8/3}$ paths and the hiding condition is equivalent to mutual avoidance. 
We look now to iterate the above calculations, motivated by the desire to deal with several 
mutually avoiding $SLE_{8/3}$ paths.

\subsection{Several $SLE_{8/3}$ paths in the half-plane}

%%%%%%%%%%%%%%%%%%%%%%%%%%%%%%%%%%%%%%%%%%%%%%%%%%%%%%%%%%%%
\begin{figure}
  \centering
  \scalebox{0.22}{\includegraphics{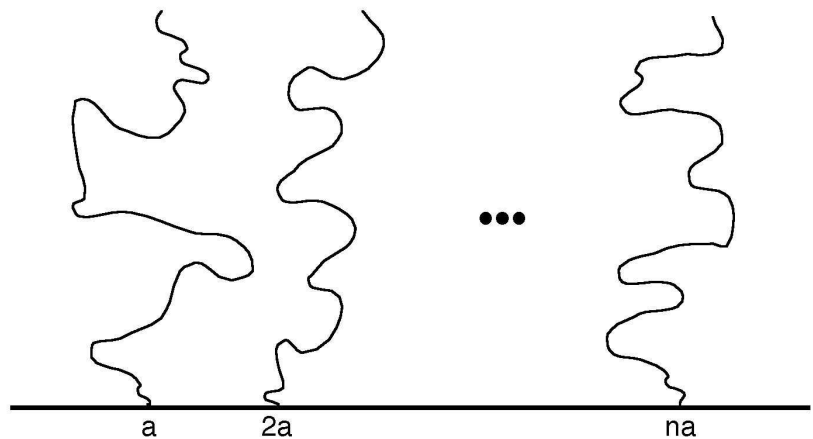}}
  \caption{Mutually avoiding $SLE_{8/3}$ paths in the half-plane.}
  \label{sle8on3}
\end{figure}
%%%%%%%%%%%%%%%%%%%%%%%%%%%%%%%%%%%%%%%%%%%%%%%%%%%%%%%%%%%%

If we now condition on the hiding event, the \SLEkp right boundary of $K_{\alpha}$ becomes an $SLE(8/3,\tilde{\rho})$ process. 
This can be viewed as the right boundary of a new one-sided restriction measure and we can in turn investigate the probability 
that this is hidden by another restriction sample to its right. 
In this way the process that gave us the hiding exponents can be iterated.  

In particular, consider $n$ independent $SLE_{8/3}$ started at points $a$, $2a$, ... , $na$ 
on the real line and conditioned not to intersect, as depicted in Figure \ref{sle8on3}.
The rightmost $SLE_{8/3}$ is an $SLE(8/3,\rho_{n})$, which is the right boundary of a one-sided 
restriction measure of exponent $\alpha_{n}$. 
To begin we have $\rho_{1}=0$ and $\alpha_{1}=5/8$. 
Furthermore from the previous results (\ref{SLEkpAVOID}) and (\ref{SLEkpREST})
\begin{eqnarray}
\rho_{n+1} 	&=&	 \frac12{\kappa} -2 + \kappa \sqrt{\frac{4\alpha_{n}}{\kappa}+\left(\frac{2}{\kappa}-\frac{1}{2}\right)^{2}},  \\
\alpha_{n} 	&=&	 \frac{1}{32}{(\rho_{n}+2)(3\rho_{n}+10)},
\end{eqnarray}
which are further simplified when we put $\kappa=8/3$. It now follows that
\begin{eqnarray}
\rho_{n} 		&	=	&	2(n-1),			\\	
\label{alphan}
\alpha_{n} 	&	=	&	 \frac18{n(3n+2)}.	
\end{eqnarray}

The final restriction exponent $\alpha_{n}$ differs from the $5n/8$ expected for $n$ independent 
(and possibly intersecting) $SLE_{8/3}$ by
\begin{equation}
\label{ss1}
\sigma		=		\frac38{n(n-1)}.
\end{equation}
We conclude that the probability that these $n$ independent $SLE_{8/3}$ are mutually avoiding scales like 
$a^{\sigma}$ as $a\rightarrow 0$. 
This corresponds to the self-avoiding walk (SAW) exponents of Duplantier and Saleur \cite{DS86} in the 
following fashion (assuming the $SLE_{8/3}$  $-$ SAW correspondence).

\begin{itemize}
	\item View $a$ as characterizing the step size for the SAWs and $N$ as the number of steps. 
	Since $SLE_{8/3}$ and hence SAW have fractal dimension $\frac43$, the probability the SAW 
	are mutually avoiding scales like $N$ raised to the power of $\frac34(3n(1-n))/8$ as $N\rightarrow \infty$.
	\item From \cite{LSW04} (using $SLE$ techniques) the number of SAWs in the half-plane scales like $N$ raised to the 
	power of $-\frac{3}{64}$ as $N\rightarrow \infty$.
	\item Therefore, the number of configurations $C_{N}$ of $n$ independent and mutually avoiding 
	self-avoiding walks scales like $N$ to the sum of these exponents, that is 
	$$C_{N}\sim N^{\frac{3n(5-6n)}{64}} \quad \mbox{as } N\rightarrow\infty .$$
\end{itemize}
This is precisely the result arrived at by Duplantier and Saleur \cite{DS86}.

%Exponents in the Wedge

\section{Wedge exponents}

\begin{figure}
  \centering
  \scalebox{0.20}{\includegraphics{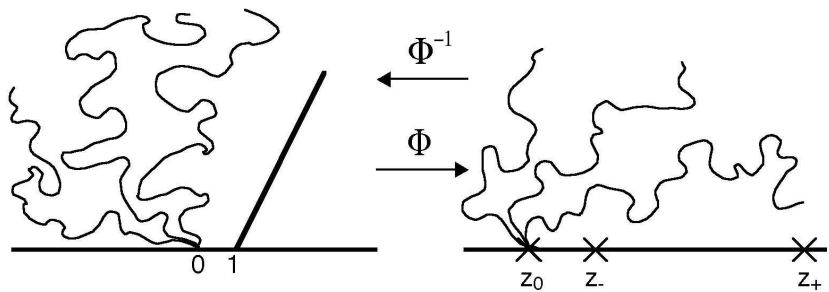}}
  \caption{The conformal map $\Phi$ removing a ray in the half-plane}
  \label{map}
\end{figure}

We now extend exponents in the half-plane to a wedge-shaped geometry with 
internal wedge angle $\theta\pi$ for $\theta\in(0,1)$.
As a byproduct of each earlier exponent calculation, 
the law of the right boundary of our collection of curves $K$ was given in terms of a one-sided restriction measure, 
let this be of exponent $\alpha$ for the time being. 
Also assume, by translating if necessary, that the right boundary begins at the origin. 
As in Figure \ref{map}, draw a ray starting at $1$ on the real line, of length $R$, and making angle $\theta\pi$ 
with the negative real line. 
The collection of curves avoids this ray if and only if its right boundary does, 
a probability which we now calculate using the restriction property. 

%%%%%%%%%%%%%%%%%%%%%%%%%%%%%%%%%%%%%%%%%%%%%%%%%%%%%%%%%%%%
%%%%%%%%%%%%%%%%%%%%%%%%%%%%%%%%%%%%%%%%%%%%%%%%%%%%%%%%%%%%

A conformal map from the half-plane to the half-plane minus the ray is 
\begin{equation}
\label{ray_map_1}
\Phi^{-1}(z) = 1 + (z-1+R\theta)^{\theta}(z-1-R(1-\theta))^{1-\theta}.
\end{equation}
{}From (\ref{ray_map_1}) it is clear that $\Phi(z)/z\rightarrow 1$ as $z\rightarrow\infty$. 
Note that $\Phi$ will also fix infinity, but not zero. 
However we can consider $\Phi(z)-\Phi(0)$ which will fix zero, infinity and scale like $z$ for large $z$. 
Then the restriction property tells us that
\begin{eqnarray}
\label{restwedge}
\mathbf{P}[K \cap ray=\emptyset] 	&	=	& 	((\Phi-\Phi(0))'(0))^{\alpha} \nonumber\\
							&	=	&	(\Phi'(0))^{\alpha}.
\end{eqnarray}
Thus all that remains is to find $\Phi'(0)$. This is easier said than done, 
since $\Phi^{-1}(z)$ as given in (\ref{ray_map_1}) is difficult to invert. 
In light of this we use the inverse function theorem to write
\begin{equation}
\label{ift}
\Phi'(0) = \frac{1}{(\Phi^{-1})'(z_{0})},
\end{equation}
where $\Phi^{-1}(z_{0})=0$. 
First consider the behaviour of $z_{0}$ for large $R$. 
{}From (\ref{ray_map_1}), $\Phi^{-1}$ extends to map both $z_{-}=1-R\theta$ and $z_{+}=1+R(1-\theta)$ to $1$.  
This implies that $z_{0}\leq z_-$. 
Writing $z_{0}$ as $z_- +kR^{-c}$ for large $R$ (and some coefficient $k$ and exponent $c$) and noting that
\begin{equation}
\label{posit}
0 = \Phi^{-1}(z_{0})	 = 1 + (z_{0}-z_-)^{\theta}(z_{0}-z_+)^{1-\theta}, 
\end{equation}
it follows that $-1 = (kR^{-c})^{\theta}(kR^{-c}-R)^{1-\theta}$.
As a consequence we can conclude that $c\geq0$, 
as otherwise the right hand side is dominated by a positive power of $R$ as $R\rightarrow\infty$. 
Since we are interested only in the scaling behaviour, assume without loss of generality that $k=1$. 
Continuing, we have $-1	= R^{-c\theta}R^{1-\theta}(R^{-1-c}-1)^{1-\theta}$.
Applying the binomial theorem to the right it is clear that $R^{-c\theta+1-\theta}$ dominates as $R$ tends to infinity. 
As the left hand side states this dominant exponent must be zero,
\begin{equation}
\label{c}
c = \frac{1-\theta}{\theta}.		
\end{equation}

Having established the behaviour of $z_{0}$ for large $R$ we now differentiate $\Phi^{-1}$ 
to find $(\Phi^{-1})'(z_{0})$. 
Note that
\begin{equation}
\label{derivative}
(\Phi^{-1})'(z) = (1-\theta)\left(\frac{z - z_-}{z-z_+}\right)^{\theta} + \theta\left(\frac{z-z_+}{z-z_-}\right)^{1-\theta}.
\end{equation}
Evaluating this at $z_{0}$ and making use of (\ref{posit}) gives
\begin{equation}
\label{derivative@z0}
(\Phi^{-1})'(z_{0}) =	-(1-\theta)(z_0 - z_+)^{-1}-\theta(z_0 - z_-)^{-1}.
\end{equation}
Applying the binomial theorem for large $R$ and arguing as above implies that 
\begin{equation}
\label{asymptotics} 
(\Phi^{-1})'(z_{0})  \sim	R^{c}  \quad \mbox{as } R\rightarrow\infty.							
\end{equation}

We now combine the restriction property, inverse function theorem and our expression (\ref{c}) for $c$ in terms of $\theta$,
the ray opening angle. 
{}From this set of calculations, the probability that our collection of curves avoids the ray scales as 
\begin{eqnarray}
\mathbf{P}[K \cap ray=\emptyset] 	&	\sim	& 	R^{-\alpha(\frac{1-\theta}{\theta})}  \quad \mbox{as } R\rightarrow\infty.
\end{eqnarray}
This computation is now used to extend the half-plane exponents to the wedge.

\subsection{Several $SLE_{8/3}$ in a wedge}

We view the probability of avoiding the ray as equivalent to the probability that $n$ $SLE_{8/3}$ of radius $R$ 
stays within the wedge of the same depth. 
To see how this probability scales with $N$, recall that the rightmost $SLE_{8/3}$ has restriction exponent 
$\alpha_{n}$ given by (\ref{alphan}) and fractal dimension $4/3$. 
Thus this probability decays like
\begin{equation}
N^{-\frac{3n(3n+2)}{32}(\frac{1}{\theta}-1)} 		\quad		\mbox{as } N\rightarrow\infty .
\end{equation}
This exponent may be added to the counting exponent in the half-plane to obtain the analogous counting exponent in the wedge, 
with $C_{N}  \sim N^{\gamma(n,\theta)}$ where 
\begin{equation}
\gamma(n,\theta) = \frac{27n}{64} - \frac{3n(3n+2)}{32\theta}.
\label{DS}
\end{equation}
This is precisely the set of exponents obtained by Duplantier and Saleur \cite{DS86}.
However, as we have used rigorous $SLE$ techniques, this derivation is complete modulo the assumption that 
$SLE_{8/3}$ is indeed the continuum limit of the self-avoiding random walk.

\subsection{Hiding exponents}

As an indication of the generality of these arguments, we now extend the new class of hiding exponents 
introduced by Werner already discussed in the half-plane. 
Again, it is a straightforward calculation. 
Return to the situation as illustrated in Figure \ref{hiding}. 
{}From (\ref{he1}) the boundary of $K_{\alpha}$ is an $SLE(8/3,\rho)$ where
\begin{equation}
\label{hew1}
\rho = \frac{1}{3}(-8+2\sqrt{1+24\alpha}).
\end{equation}
Now using (\ref{SLEkpAVOID}) and the above we can condition the boundary to hide another restriction 
measure of exponent $K_{\beta}$ which makes it an $SLE(8/3,\tilde{\rho})$ where
\begin{equation}
\label{hew2}
\tilde{\rho} = -\frac{2}{3} + \frac{8}{3}\sqrt{\frac{3}{2}\beta+\left(\frac{1}{4}\sqrt{1+24\alpha}-\frac{3}{4}\right)^{2}}.
\end{equation}
Turning to (\ref{SLEkpREST}) this $SLE(8/3,\tilde{\rho})$ may be viewed as a sample of one-sided restriction measure of exponent
\begin{equation}
\tilde{\alpha} = \beta +\frac{1}{3} + \frac{1}{24}(\sqrt{1+24\alpha}-3)^{2}
+\sqrt{\frac{3}{2}\beta+\frac{1}{16}\left(\sqrt{1+24\alpha}-3\right)^{2}}.
\end{equation}
It follows that the probability that the two restriction samples stay inside the wedge will scale like
\begin{equation}
R^{-\tilde{\alpha}(\frac{1}{\theta}-1)} 	\quad		\mbox{as } R\rightarrow\infty .
\end{equation}
We therefore conclude that the hiding exponent in the wedge is simply $\sigma-\tilde{\alpha}(1/\theta-1)$, 
where $\sigma$ is the corresponding exponent in the half-plane. 
This simple procedure can be extended to all exponents described with \cite{Wer04}, 
extending each result to wedge geometries.

\section{Conclusion}

The $SLE$ formalism is known to provide an ideal framework in which to investigate the properties of various random curves. 
When coupled with the restriction property and absolute continuity relations governing $SLE(\kappa,\rho)$,
an iterative approach allows easy exploration of several mutually avoiding interfaces. 
Indeed, as shown, a wealth of exponents for the self-avoiding random walk, a notoriously difficult problem, 
can be established modulo the assumption that $SLE_{8/3}$ is the continuum limit for the self-avoiding walk. 
Although making this assumption may seem to detract from the otherwise rigorous nature of $SLE$, 
the potential importance of $SLE$ simply as a calculational tool should not be neglected. 
It was with considerable ingenuity that so many exponents for the self-avoiding walk were able to be 
established  using general arguments combined with scaling dimensions obtained using Coulomb gas 
and later Bethe Ansatz techniques (see for example \cite{Car84,CR84,DS86,BS93,BBO98}).
The ease with which some of these exponents follow from the $SLE$ approach is not to be taken for granted.

An additional benefit of $SLE$, as also illustrated in this paper, is its ability to provide new results, 
as well as confirming older ones. 
The hiding exponents first introduced by Werner \cite{Wer04} have been extended to wedge geometries, 
and more generally this paper indicates how the iterative process first outlined by Werner may be 
coupled with the restriction property.
This provides exponents for the joint behaviour of several restriction 
measures in geometries contained within the half-plane. 
Special cases such as restriction seem crucial to generalising the powerful tools of the $SLE$ project to multiple 
$SLEs$ or multiply connected domains where \SLEk lacks a natural definition \cite{Car07}. 

\ack 

This work has been partially supported by the Australian Research Council.

\end{document}